# Masses of Stellar Black Holes and Testing Theories of Gravitation

K. A. Postnov and A. M. Cherepashchuk

*Sternberg Astronomical Institute, Moscow State University, Universitetskiĭ pr. 13, Moscow, 119899 Russia*

**Abstract**—The paper analyzes the mass distribution of stellar black holes derived from the light and radial-velocity curves of optical stars in close binary systems using dynamical methods. The systematic errors inherent in this approach are discussed. These are associated primarily with uncertainties in models for the contribution from gaseous structures to the optical brightness of the systems under consideration. The mass distribution is nearly flat in the range 4–15 $M_\odot$. This is compared with the mass distribution for black holes in massive close binaries, which can be manifest as ultrabright X-ray sources ($L_x > 10^{39}$ erg/s) observed in other galaxies. If the X-ray luminosities of these objects correspond to the Eddington limit, the black-hole mass distribution should be described by a power law, which is incompatible with the flat shape derived dynamically from observations of close binaries in our Galaxy. One possible explanation of this discrepancy is the rapid evaporation of stellar-mass black holes predicted in recent multi-dimensional models of gravity. This hypothesis can be verified by refining the stellar black-hole mass spectrum or finding isolated or binary black holes with masses below $\sim 3 M_\odot$.

## 1. INTRODUCTION

The discovery of approximately twenty black holes with stellar masses and about one hundred supermassive black holes (e.g. reviews [1, 2]) raises the question of their demography; i.e., the relationship of these extreme objects to other objects in the Universe, as well as to the deep physical properties of space–time. Bimodality in the mass distribution for stellar-mass relativistic objects was recently detected [3–5]. The masses of neutron stars lie in the narrow range $M_{NS} = (1-2) M_\odot$, with the mean value being $(1.35 \pm 0.15) M_\odot$. On the other hand, the masses of black holes are distributed over the fairly broad range $M = (4-15) M_\odot$, with the mean value being $8-10 M_\odot$. No neutron stars or black holes have been found in the mass interval $2-4 M_\odot$, despite the fact that masses have now been measured for almost 40 relativistic objects. This gap in the mass distribution of relativistic objects at masses of $2-4 M_\odot$ cannot be due to observational selection effects [2–5], and seems especially surprising from the viewpoint of recent data on the mass distribution of the CO cores of Wolf–Rayet stars at the end of their evolution [5], which cover the wide range $M^f_{CO} = (1-2)-(20-44) M_\odot$ and are distributed continuously. Since Wolf–Rayet stars in close binary systems are commonly thought to be the progenitors of relativistic objects [6–8], the large difference between the distribution of the final masses of the CO cores of Wolf–Rayet stars and the masses of the resulting relativistic objects is a non-trivial observational fact, which must be explained.

The bimodal mass distribution for relativistic objects was interpreted in [9, 10] in terms of modern concepts about the late stages of stellar evolution and explosions of collapsing supernovae (Types II and Ib, c). As an alternative way to explain the broad distribution of masses for stellar-mass black holes and the lower limit observed in binary systems, $\sim 4 M_\odot$, we consider here modern, multi-dimensional theories of gravity, which enable us to view the mechanism and characteristic time for the quantum evaporation of black holes in a new light.

## 2. METHODS FOR DETERMINING BLACK-HOLE MASSES IN BINARY SYSTEMS

It is important to answer the question of whether the observed broad distribution of black-hole masses in the interval $4-15 M_\odot$ is real, or whether these masses are actually distributed in accordance with some other law (for example, concentrated in a narrower interval), with the observed scatter being due to errors in the derived masses.

Most of the information about the mass of a black hole in a binary system in which the secondary is an optical star is contained in the mass function of the optical star [2, 11]:

$$f_v(m) = \frac{m_x^3 \sin^3 i}{(m_x + m_v)^2} \quad (1)$$
$$= 1.038 \times 10^{-7} K_v^3 P (1-e^2)^{3/2},$$

which is derived from its radial-velocity curve; the optical star is considered to be a point mass moving along a Keplerian ellipse. Here, $m_x$ and $m_v$ are the masses of the black hole and optical star in solar units, $K_v$ the semi-amplitude of the radial-velocity curve of the optical star (in km/s), $P$ the orbital period of the binary system (in days), and $e$ the eccentricity of its orbit. In reality, the optical star is not a point mass, since its shape is disturbed by tidal interactions with the black hole and its atmosphere is heated by X-ray radiation from the black-hole accretion disk. Taking these effects into account shows that they affect the derived black-hole mass most when the component-mass ratio is $q = m_x/m_v < 1$ [11]. In this case, the center of mass of the binary system is located inside the optical star (as occurs, for example, in the systems Cyg X-1, LMC X-1, and SS 433, in which $q = 0.3–0.6$), and the distortion of the spectral-line profiles used to derive the radial-velocity curve is greatest. When $q = 0.3–0.6$, corrections of the mass function $f_v(m)$ for the effects of the finite size of the optical star do not exceed 10%, and can be reliably estimated using modern methods for synthesizing the line profiles and radial-velocity curves of X-ray binaries [11–13].

In the case of X-ray binaries with massive (O–B) companions (Cyg X-1, LMC X-1, LMC X-3, and SS 433), there is another effect that disturbs the line profiles and radial-velocity curve of the optical star: the variable (depending on the phase of the orbital period) selective absorption of the light of the optical star by its intense stellar wind. (The mass-loss rates of such stars are typically $\sim 10^{-6}–10^{-7} M_\odot/$yr, and reach $10^{-4} M_\odot/$yr in the case of SS 433.) The absorption coefficient at the line center is considerably greater than in the neighboring continuum. Therefore, the central part of the absorption line is formed in the upper layers of the stellar atmosphere, at the base of the stellar wind, where the radial velocity of the plasma outflow reaches a few tens of km/s. Since the free-fall acceleration in a star with an almost filled Roche lobe varies over the stellar surface, the velocity and intensity of the wind near its base will also vary over the stellar surface, resulting in additional orbital-phase-dependent Doppler shifts of the absorption lines in the spectrum of the optical star and distortion of its radial-velocity curve [14]. Moreover, in the case of an X-ray binary with an elliptical orbit, nonradial pulsations can be excited in the optical star, as occurs in the system containing the neutron star Vela X-1 [15]. This also results in additional distortion of the radial-velocity curve of the optical star and leads to systematic errors in the derived mass of the relativistic object.

In the case of large mass ratios $q > 1$, the center of mass of the system is located outside the body of the optical star, and the effect of the finite size of the optical star becomes small. This is especially important because the masses of 15 of 18 black holes were determined in transient, low-mass X-ray binaries (X-ray novae with large mass ratios, $q > 1.5$). Therefore, the masses of most of the black holes are affected only slightly by the finite sizes of their optical components. Since the stellar winds from the low-mass (A–M) stars that are the companions of the black holes in X-ray novae are weak, the effect of selective absorption of the light of the optical stars by their wind is also small. The orbits of all X-ray novae with low-mass (A–M) companions are circular, and the optical stars in these systems fill their Roche lobes.

The mass of the invisible companion (black hole) in a binary system is derived from the mass function of the optical star $f_v(m)$:

$$m_x = f_v(m)\ \left(1 + \frac{1}{q}\right)^2 \frac{1}{\sin^3 i}. \qquad (2)$$

The uncertainty in the black-hole mass includes random and systematic errors. The random errors can be reduced by increasing the accuracy and duration of the observations. The systematic errors are due to uncertainty in the model for the X-ray binary. Taking the systematic errors into account when determining black-hole masses is very difficult. Let us consider the influence of systematic errors in the parameters $q, i$ on the corresponding estimate of the black-hole mass $m_x$.

The parameter $q$ is usually estimated from the rotational broadening of absorption lines in the spectrum of the optical star. In most close X-ray binaries containing black holes (in particular, in X-ray novae), the optical star fills its Roche lobe, whose relative size depends on the mass ratio $q$. On the other hand, the larger the absolute size of the optical star, the greater the rotational broadening of absorption lines in its spectrum. As a result, assuming the axial and orbital rotations are synchronized, we obtain the following equation determining $q$ [1, 11–13]:

$$v \sin i = 0.462 K\ _v q^{-1/3} \left(1 + \frac{1}{q}\right)^{2/3}. \qquad (3)$$

The rotational broadening $v \sin i$ varies with the phase of the orbital period, since the dimensions of the star along the line connecting the component centers are different from those perpendicular to this direction [11]. In addition, X-ray heating of the optical star gives rise to an emission component in the lines that depends on the phase of the orbital period, which distorts the standard absorption-line profile [13]. The corresponding errors in $v \sin i$ values derived from the analysis of absorption-line profiles in the spectrum of the optical star can be as large as 10–20%. A new method for determining $q$ and $i$ from the orbital variability of the absorption-line profiles in the optical spectra of close X-ray binaries was suggested in [16, 17]. Modern methods for synthesizing line profiles in the spectra of the optical components of close X-ray binaries can take into account the spectral variability of the optical star and thereby reduce the systematic errors as much as possible. We emphasize that most black-hole masses have been measured using X-ray novae, for which $q > 1$, in the quiescent state. In this case, the effect of X-ray heating is small, and the error in $q$ affects the value of $m_x$ when $q > 1$ only weakly [see (2)]. As a result, the influence of systematic errors on $q$ values derived from the rotational broadening of absorption lines is usually insignificant.

The orbital inclination $i$ is most affected by systematic errors. A method for determining $i$ from the optical light curves of X-ray binaries, whose shapes are determined primarily by the ellipticity of the optical star, was proposed in [18, 19], and is now being widely used to derive the masses of black holes in binary systems [1, 11]. The main source of systematic errors in the $i$ values derived using this method is the contribution of gaseous structures (such as the accretion disk, gas jets, and the region of interaction between the jet and disk) to the total optical or infrared luminosity of the system. This contribution can be estimated spectrophotometrically by comparing the equivalent widths of absorption lines in the spectrum of the binary system with the corresponding equivalent widths in the spectrum of an isolated star of the same spectral and luminosity class. Unfortunately, the contribution of such gaseous structures can exceed 50% in the case of X-ray novae—binaries with low-mass cool stars [20], and the orbital variability of the emission of these structures is complex [21]. As a result, the systematic errors in $i$, and consequently in the mass $m_x$, become considerable. For example, the mass of the black hole $m_x$ in the X-ray nova GRO J0422+32 estimated using two different methods varies from $2.5-4M_\odot$ to $>9M_\odot$ [20]. In the case of quasi-stationary close X-ray binaries with massive, hot stars (Cyg X-1, LMC X-1, and LMC X-3), the contribution of the optical radiation of gaseous structures is small ($<2\%$), but the optical light curves of these systems suffer from the effects of absorption of the light from the optical star by the gaseous structures [22], which also leads to systematic errors in the orbital inclinations of these close binaries. In addition, the optical stars in the Cyg X-1 and LMC X-1 systems do not quite entirely fill their Roche lobes. This also introduces an extra systematic error into $i$, so that information about the distance to the system is required [11].

The new method for determining $q$ and $i$ presented in [16, 17], based on analysis of the orbital variability of absorption-line profiles in the spectrum of the optical star in a close X-ray binary, does not depend on the contribution of gaseous structures to the total luminosity of the system. Therefore, high-quality spectroscopy of close X-ray binaries with high spectral resolution $R = 50\,000-100\,000$ using the largest new-generation telescopes should enable us to appreciably reduce the effect of systematic errors and obtain the most trustworthy estimates of black-hole masses in close binary systems.

Another opportunity for independently determining the orbital inclinations $i$ of X-ray binaries is the use of new, more accurate information on their distances $d$, which will be provided by next-generation astrometric space observatories (SIMA, GAIA, etc.). Knowledge of the distance $d$, interstellar absorption $A_v$, apparent magnitude $m_v$, and the contribution of gaseous structures to the system's luminosity enables determination of the average radius of the optical star $R_v$. This gives us a relation between $q, \mu$, and $i$ [11, 23] (where $\mu$ is the degree of filling of the Roche lobe by the optical star):

$$\sin i = \left(\frac{0.38\mu}{R_v}\right) \left(\frac{GP^2 f_v(m)}{4\pi^2}\right)^{1/3} \left(\frac{1+q}{q^{1.208}}\right). \quad (4)$$

Since we can assume for X-ray novae that $\mu = 1$ [2] and the value of $q$ can be determined independently from the rotational broadening of absorption lines in the optical spectrum [see (3)], Eq. (4) can be used to obtain an independent determination of the orbital inclination $i$.

Thus, the probable presence of considerable systematic errors in dynamical black-hole masses currently prevents us from firmly establishing the black-hole mass distribution. We shall consider below only two limiting cases of the distribution of black-hole masses in close binaries: (1) a sharp $\delta$-function-like peak near some specified value $\sim 9-10 M_\odot$ and (2) a uniform distribution over a broad range ($4-15 M_\odot$).

## 3. DIFFERENCES IN THE OBSERVABLE MANIFESTATIONS OF ACCRETING NEUTRON STARS AND BLACK HOLES

Since it will be important for us that the masses of black holes and of neutron stars have different lower limits, it is appropriate to underline here the fundamental differences in the observable manifestations of these two types of relativistic objects. As was noted above, the masses of approximately 40 compact objects in binary systems—18 black holes and more than 20 neutron stars—have currently been measured. It is a remarkable fact that the observed features of the accreting neutron stars and black holes differ, in accordance with the quantitative predictions of Einsteinian general relativity: there is a gap in the observed manifestations of compact relativistic objects near a mass of $3M_\odot$ (the absolute upper limit for the mass of a neutron star following from general relativity). In all cases when the mass of an X-ray pulsar, type I X-ray burster, or radio pulsar (phenomena showing clear signs that we are observing the surface of a relativistic compact object), the corresponding masses do not exceed $3M_\odot$, in good agreement with general relativity. The large number of known neutron-star masses (over 20) makes this result statistically significant. On the other hand, nome of the 18 massive ($> 3M_\odot$) relativistic objects (black-hole candidates) is associated with an X-ray pulsar, type I X-ray burster, or radio pulsar. Therefore, none of the black-hole candidates shows features associated with an observable surface, as should be the case for black holes in general relativity. The increasing number of such objects (currently 18) confirms the high statistical significance of this result. Of course, the absence of clear features associated with an observable surface represents a necessary, but not sufficient, indication of a black hole.

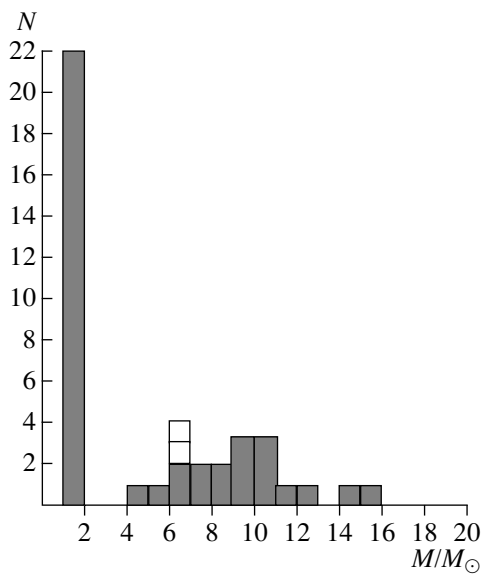

**Fig. 1.** Observed mass distribution of relativistic objects in close binary systems. The neutron stars are concentrated in the narrow interval $M = (1-2)M_\odot$, while black holes in close binaries are found in the interval $4-16M_\odot$. The masses of isolated black holes derived from microlensing observations are marked by hollow squares.

We emphasize that there are also finer observational spectral differences between accreting neutron stars and black holes, as well as differences in the rapid variability of their X-ray emission (see, for example, [24, 25]). These differences are also consistent with the idea that neutron stars have observable surfaces, while black holes do not.

Therefore, all the necessary conditions imposed by general relativity on the observable manifestations of accreting neutron stars and black holes are satisfied. This strengthens our certainty in the existence of black holes in nature. Further, there is hope that sufficient criteria for the observational identification of black holes will be obtained very soon using the X-ray space interferometer [26] and via the detection of bursts of gravitational radiation due to merging black holes in binaries using gravitational-wave interferometers such as LIGO, VIRGO, and LISA (for more details, see the review [27]).

## 4. THE OBSERVED BLACK-HOLE MASS FUNCTION

Thus, modern astronomical data provide a basis for a discussion of the observed mass function for stellar-mass black holes (Fig. 1). The apparent mass distribution is in the range $\sim 4$ to $\sim 20M_\odot$, with no significant concentration at any particular mass. Since the systematic errors in the mass of an invisible companion in a close binary (especially ambiguity in the orbital inclination) can appreciably distort the true distribution, we shall consider two limiting cases: (a) a narrow distribution of masses around some specified value $M_0$ (for definiteness, we adopt $M_0 = 10M_\odot$) and (b) a uniform distribution over some range $M_{\min} - M_{\max}$, where $M_{\min} = (3-4)M_\odot$ and $M_{\max} = (15-20)M_\odot$.

*Case (a):* a narrow mass function around a specified value $M_0$, $dN/dM \sim \delta(M - M_0)$. There is no fundamental physical justification for the realization of such a distribution. Moreover, the observed X-ray luminosity function of massive close binary systems in other galaxies is clearly inconsistent with this hypothesis (see below).

*Case (b):* a flat (or almost flat) mass function $dN/dM \sim M^{-\beta}$, where $\beta \approx 0$, $M_{\min} < M < M_{\max}$. Such a distribution seems more probable, since the masses of collapsing supernova progenitors are distributed over some interval, and the fraction of the stellar mass that collapses into the black hole can depend substantially on the physical conditions of the collapse (such as rotation, magnetic field, etc.). Accordingly, we will consider this case to be realized below.

### 4.1. Initial Black-Hole Mass Function: Direct Calculations

The mass function is a fundamental characteristic of black holes. Modern theoretical concepts about the collapse of stellar cores are incomplete, and cannot predict unambiguously the masses of the resulting compact remnants. For example, a bimodal initial mass function (IMF) for the compact remnants (with peaks at $M_{NS} = 1.28 M_\odot$ for neutron stars and $M_{BH} = 1.78 M_\odot$ for black holes) was obtained in calculations of collapsing type II supernovae [28], in clear contradiction with the absence of observed black-hole candidates with masses below $3$–$4 M_\odot$. In contrast, based on certain assumptions about the masses of the resulting black holes, Fryer and Kalogera [10] obtained a broad, continuous distribution of black-hole masses up to $10$–$15 M_\odot$, without any deficit of objects with masses $1.5$–$3 M_\odot$. Neither of these theoretical distributions are in satisfactory agreement with the observations. The origin of these discrepancies may lie in observational selection effects. For example, Fryer and Kalogera [10] suppose that, if a black hole acquires some additional velocity ("kick") during its formation, low-mass black holes in binary systems will have a lower probability of surviving after the ejection of the massive envelope. This argument is doubtful, since this effect should be even stronger during the formation of neutron stasr in binary systems [5], in evident contradiction with the observed pattern.

A possible physical explanation for the absence of observed masses of compact objects in the range $1.5$–$3 M_\odot$ was proposed in [9], which considers a magnetorotational mechanism for supernova explosions [29] and a fairly soft equation of state for neutron stars, with a limiting mass of $M_{\max} \approx 1.6 M_\odot$.

However, there is no doubt that all such calculations are model-dependent, and, moreover, do not adequately take into account the effects of rotation, the magnetic field, the possible accretion of matter from the ejected envelope, and so on. It is likely that the derivation of the black-hole IMF will require the use of phenomenological data on the core masses and other physical characteristics of supernova progenitors derived from observations [5].

Nevertheless, it seems useful to analyze various hypotheses about the black-hole IMF and compare the results with observations. It is well known that the stellar IMF has the power-law form $f(M)_i = (dN/dM)_i \propto M^{-\alpha_i}$. The Salpeter IMF has an index (slope of the differential mass function) $\alpha_i = 2.35$ for stars with masses up to $10 M_\odot$ in the solar neighborhood, and is in agreement with modern observations. The slope of the IMF for more massive stars becomes steeper. (This should probably be treated only as a trend due to the large errors in the masses of early-type stars and insufficient statistics.) For example, the Miller–Scalo IMF yields $\alpha_i = 2.5$ for stars with $M \sim 10 M_\odot$. Some astronomers (e.g. B. Elmegreen, et al.) believe that the stellar IMF is a manifestation of the universal character of star formation in the turbulent self-gravitating interstellar medium in galaxies (see the recent review [30] and references therein). In addition, it is well known that the stellar winds of massive OB stars carry away a substantial fraction of the initial mass of the main-sequence stars, and the masses of supernova-progenitor cores are distributed over a wide range [5]. It is easy to see that power-law dependences for the fundamental parameters of stars (such as their luminosity and radius) on their mass can lead to a stellar mass distribution at the end of their thermonuclear evolution (just before the collapse) that also has a power-law form. Therefore, the power-law form of the black-hole IMF is admissible theoretically, but does not follow from any general physical arguments.

### 4.2. Variations in the Black-Hole Mass in the Course of Subsequent Evolution

The mass of a black hole that has formed in any way can either (a) increase due to accretion of matter (or, more precisely, energy) onto the black hole or (b) decrease due to quantum evaporation [31]. The mass $M$ of an isolated black hole moving with speed $v$ through an interstellar medium with density $\rho$ and sound speed $v_s$ increasess due to Bondi–Hoyle accretion $\dot{M}^+ \propto \rho v M^2/(v_s^2 + v^2)^2$. In the typical case, $v_s < 1$ km/s, and the velocity dispersion of massive stars in the Galactic disk (which can give birth to black holes at the end of their evolution) is of the order of 10 km/s, so that $\dot{M}^+ \sim \rho M^2/v^3 \sim 10^{13}$ g/s $= 10^{-13} M_\odot$/yr for an isolated black hole with a mass of a few solar masses moving through a medium with a characteristic density of $10^{-23}$ g/cm$^3$. Therefore, the increase in the masses of isolated black holes in the Galaxy can be neglected. A unique opportunity to measure the masses of isolated black holes via observations of gravitational microlensing has recently appeared [32, 33]. This method has enabled the measurement of the masses of two black-hole candidates using the microlensing events MACHO-96-BLG-5 ($M = 6^{+10}_{-3} M_\odot$) and MACHO-98-BLG-6 ($M = 6^{+7}_{-3} M_\odot$) [32]. The corresponding values are also plotted in Fig. 1 (hollow squares).

The mass of a black hole in a close binary system can increase due to the accretion of matter from its companion. In the case of low-mass binaries containing black holes (such as X-ray novae), the average accretion rate is determined by the evolution of the binary orbit as its orbital angular momentum is carried away by gravitational radiation or the magnetized stellar wind of the optical star, and will be of the order of $\dot{M}^+ \sim (10^{-9}$–$10^{-10}) M_\odot$/yr. The accretion rate can be greater for black holes in massive close binaries (such as Cyg X-1 and SS 433). If the standard regime of accretion in a thin disk is realized [34, 35], the rate of increase of the black-hole mass will be limited by the Eddington luminosity (about $10^{-7}$ $M_\odot$/yr for $M = 10 M_\odot$). In the case of advection-dominated accretion, the rate of increase in the mass can be even greater. However evolutionary considerations indicate that there should be substantially fewer accreting black holes in massive close binary systems in our Galaxy [36]. The increase in the mass of a black hole will obviously be determined by the duration of the accretion stage ($\sim (10^7-10^8)$ yr for low-mass and $\sim 10^5$ yr for massive close binaries). Therefore, in the case of black holes in close binaries, we can neglect to first approximation the possible increase in their mass by about 10%.

### 4.3. The Black-Hole Mass Function and the Luminosity Function of X-Ray Sources in Galaxies

The high angular resolution of the modern CHANDRA and XMM X-ray telescopes makes it possible to study individual X-ray sources in other galaxies and, in particular, to construct their distribution over the observed X-ray luminosity; see, for example, [37], as well as [38], which presents the X-ray luminosity function constructed using the HRI instrument onboard the ROSAT satellite. These and other works (see also, for example, the recent review [39]) have shown that the luminosity function of point-like X-ray sources in various galaxies has a power-law form $dN/dL_x \propto L_x^{-\beta}$ over a wide range of luminosities $10^{36}$ to $\sim 10^{40}$ erg/s, with the index being $\beta \sim 1.5-1.7$. The hypothesis that there exists a universal X-ray luminosity function with index $\beta \approx 1.6$ for the population of binary systems in galaxies was put forward and argued in [37]. As was shown in [40], the existence of a universal power-law for this X-ray luminosity function can be explained by the nature of accretion onto compact objects in massive close binary systems. The characteristic properties of the observed X-ray luminosity function are (1) the absence of a visible break at $L_x \approx 10^{38}$ erg/s (the Eddington limit for accretion onto a neutron star) and (2) a sharp cutoff in the function at a luminosity of $\sim (2 \times 10^{39} - 2 \times 10^{40})$ erg/s. Although the corresponding observations may be statistically incomplete, let us consider what we can deduce about the masses of accreting black holes in binary systems based on the X-ray luminosity function.

Let us begin with the cutoff of the observed luminosity at $\sim 2 \times 10^{40}$ erg/s. Let us suppose that this maximum luminosity is equal to the Eddington luminosity, $L_{Edd} \approx 10^{38}(M/M_\odot)$ erg/s. Depending on its inclination, the luminosity of a standard accretion disk can be a factor of three to six higher than the nominal Eddington luminosity (see discussion in [37]). The maximum mass of the corresponding black holes would then be $M_{\max} \sim (20-30) M_\odot$. We believe that the observation of such bright X-ray sources in many galaxies is difficult to reconcile with the hypothesis that the black-hole masses are concentrated near the value $\sim 10 M_\odot$, suggesting a fairly broad distribution of black-hole masses is more likely. An alternative explanation for the ultrabright X-ray sources observed in other galaxies is that they are microquasars whose jets are directed toward the observer (see the discussion in the review [41] and references therein). In this case, the true X-ray luminosity of the source should be a factor of at least $1 - \cos\theta$ lower than the luminosity derived from the received radiation flux assuming spherical symmetry of the source (where $\theta$ is the opening angle of the collimation cone of the radiation). The estimates of [42] show that this hypothesis requires unreasonably broad collimation of the radiation $\theta \sim 30-60°$ in order to obtain agreement with the statistics of the observed ultrabright X-ray sources. In addition, the microquasar hypothesis is not consistent (at present) with the observed absence of a break in the X-ray luminosity function near $\sim 10^{38}$ erg/s.

The analysis of the X-ray luminosity function presented in Fig. 5 of [37] shows that the absolute value of the index characterizing the slope of the function $dN/dL_x$ becomes greater than the mean value $-1.6$ at a luminosity of $\sim 2 \times 10^{39}$ erg/s, namely, $dN/dL_x \propto L_x^{-2...-2.2}$. The following two conclusions can be drawn from this fact. First, including the factor of three to six noted above when interpreting the observed luminosity of an accretion disk radiating at the Eddington limit, a luminosity of $2 \times 10^{39}$ erg/s corresponds to a black-hole mass of $3-4 M_\odot$. Second, if we assume that all ultrabright X-ray sources with $L_x > 2 \times 10^{39}$ erg/s are actually close binary systems with black holes whose luminosities are about equal to the Eddington luminosity, then $dN/dL_x \propto dN/dM$, and the observed slope of the X-ray luminosity function at high luminosities should directly reflect the distribution of black-hole masses in close binaries: $dN/dM \sim M^{-2...-2.2}$. Since the increase in the masses of black holes in massive close binaries (which seem to correspond to ultrabright X-ray sources) is small during the accretion stage, the corresponding distribution should reflect the initial form of the black-hole mass function: $f_0(M) \propto M^{-2...-2.2}$.

Finally, if accretion onto a black hole in a close binary occurs in a subcritical regime, the X-ray lu-minosity for standard disk accretion is simply $L_x = \eta \dot{M} c^2$, where the coefficient of proportionality depends on the rotation of the black-hole ($\eta \approx 0.06$ for nonrotating and 0.42 for maximally rotating black holes). In this case, the luminosity function of such X-ray sources *does* not depend on the mass of the black hole and is determined, as in the case of neutron stars, by the mass distribution of the optical components in the binaries and the dependence of the accretion rate on these masses [40]. This makes the absence of a break at the value of $L_{Edd}$ for $1-2\ M_\odot$ quite natural. A cutoff in the luminosity function is expected at higher luminosities, determined by the limiting (Eddington) X-ray luminosity for the black hole with minimum mass. If $M_{\min} = 3-4 M_\odot$, the corresponding value could be a few times $10^{39}$ erg/s. The broad distribution of black-hole masses in close X-ray binaries is a supplementary factor that smooths the sharp break.

Therefore, we have arrived at two important conclusions: the distribution of black-hole masses in binary systems, $dN/dM \sim M^{-2...-2.2}$, derived from the observed X-ray luminosity function for ultrabright X-ray sources with $L_x > 2 \times 10^{39}$ erg/s in other galaxies (a) is consistent with the black-hole mass range $4-20\ M_\odot$ obtained from dynamical measurements and (b) is not consistent with a uniform distribution for the dynamical masses of black holes in close binaries within this same range. This latter conclusion can be explained by various selective effects, such as the possibility that the evolution of massive (ultrabright X-ray sources) and low-mass (most close X-ray binaries with known masses for their black-hole candidates) X-ray binaries proceeds along different paths. However, we can also seek a physical origin for the observed discrepancy that is not related to evolutionary processes. With this aim in view, let us consider the hypothesis of enhanced evaporation of stellar-mass black holes.

## 5. ENHANCED EVAPORATION OF BLACK HOLES IN CERTAIN MODERN MODELS OF GRAVITY

In the framework of a classical, four-dimensional, Einsteinian theory of gravity, the quantum evaporation of stellar-mass black holes is negligible, since the characteristic time for Hawking evaporation, which is of the order of $\tau \sim t_{Pl}(M/m_{Pl})^3$ (where $t_{Pl} \sim 10^{-43}$ s and $m_{Pl} \sim 10^{-5}$ g are the fundamental Planck time and mass), becomes shorter than the current age of the Universe $t_H \sim 14 \times 10^9$ yr only for objects with masses below $\sim 10^{15}$ g (a detailed consideration of black-hole evaporation in the framework of general relativity can be found, for example, in [43]). Consequently, if we neglect quantum evaporation, the observed spectrum of black-hole masses in close binary systems should reflect the initial mass function of black holes in these systems. From this point of view, the observed flat spectrum over a broad range of masses leads us to the conclusion [5] that not only the mass of the supernova progenitor, but also a number of other physical parameters (rotation, magnetic field, etc.), determine the mass of the black hole formed during the collapse of the stellar core.

Modern attempts to devise a unified theory of physical interactions have primarily promoted superstring theory as the most promising possibility (see the review [44]). This is considered to be the most realistic version of a quantum theory of gravity (which must describe, in particular, the evaporation of black holes). The concepts of superstring theory, always formulated in a multi-dimensional space, has led recently to multi-dimensional models of gravity with a macroscopic additional dimension (see the review [45]). Roughly speaking, these models can be subdivided into two broad classes: models with a factorized geometry (of the ADD type [46]) and models with a nonfactorized geometry (of the RS type [47]). The latter are preferable from the viewpoint of modern cosmology [45], and we shall accordingly consider black holes within the RS approach. In the simplest versions of this model, the observable physical world (i.e., particles and fields apart from gravity) is localized on a four-dimensional surface (the so-called brane) imbedded into an extra dimension (the so-called bulk) whose geometry is described by an anti-deSitter (AdS) metric. The four-dimensional metric described by classical general relativity is induced on the RS brane. The characteristic scale of the additional dimension (warp factor) is just the inverse of the radius of curvature $\mathcal{L}$ of the five-dimensional AdS metric. An extremely important (and, probably, the most fundamental) property discussed in recent years is the correspondence between supergravity in a five-dimensional AdS space and conformal field theory (supersymmetric Yang–Mills theory) on a four-brane (the so-called AdS/CFT correspondence; for more details, see the review [48] and references therein).

Attempts to derive static black-hole-type solutions within RS models have thus far been unsuccessful: "black cigar" type solutions (a black hole on a four-brane that asymptotically transforms into the AdS space [49]) are unstable [50], and clearly unable to describe the result of the collapse of a massive stellar core on the brane. There have been some attempts to obtain numerical solutions for black holes localized on a four-dimensional RS brane, but hints of a static solution were obtained numerically only for black holes whose horizons were less than $\mathcal{L}$ in size [51].

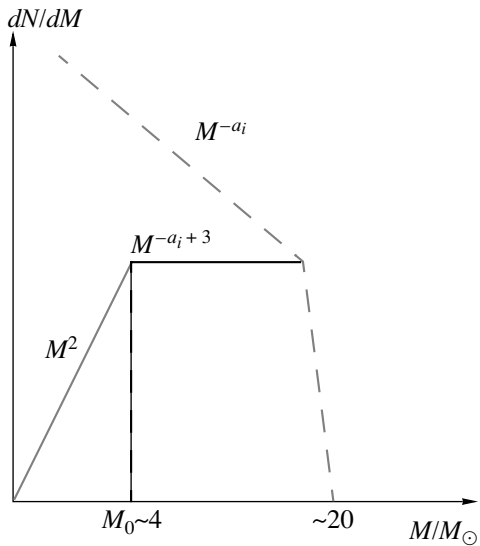

**Fig. 2.** Qualitative shape of the expected stationary black-hole mass distribution (solid curve) with a power-law initial form $(dN/dM)_i \sim M^{-a_i}$ (dashed curve) in a model with enhanced evaporation of the black hole on the RS2 brane. The mass $M_0$ corresponds to the minimum mass of a black hole that can evaporate over the Hubble time.

Analysis of the classical evaporation of black holes on the RS brane [53] shows that, if the AdS/CFT correspondence is valid for black holes, astrophysically interesting features of their evaporation will appear. Namely, the classical evaporation of four-dimensional black holes can occur much more rapidly (at least, as long as the radius of the horizon of the four-dimensional black hole on the brane is greater than $\mathcal{L}$). The evaporation time in this model is

$$\tau \equiv \frac{M}{\dot M} \sim \frac{1}{\mathcal{L}^2}(G_4 M)^3 \qquad (5)$$
$$\sim 1\,\mathrm{yr} \left(\frac{M}{M_\odot}\right)^3 \left(\frac{1\,\mathrm{mm}}{\mathcal{L}}\right)^2,$$

where $G_4$ is the effective Newtonian gravitational constant on the brane. An independent field-theoretical analysis [53], also based on the AdS/CFT correspondence, resulted in a qualitatively similar expression for the evaporation time:

$$\tau \simeq 10^2\,\mathrm{yr} \left(\frac{M}{M_\odot}\right)^3 \left(\frac{1\,\mathrm{mm}}{\mathcal{L}}\right)^2. \qquad (6)$$

The physical reason for the increase in the rate of evaporation of black holes in these models is that the evaporation rate increases in proportion to the number of degrees of freedom of the corresponding four-dimensional conformal field theory on the brane: $\propto (\mathcal{L}/l_{Pl})^2$, where $l_{Pl} \approx 10^{-33}$ cm is the classical Planck length. The discrepancy in the coefficients in the above formulas is due to the model allowance for the number of degrees of freedom in [53].

Note that the evaporation of a black hole into CFT modes produces essentially low-energy Kaluza–Klein gravitons, which are weakly coupled to the fields of ordinary matter, and are therefore unobservable by direct astrophysical methods. Moreover, the accelerated evaporation of black holes no longer takes place when the radius of the causality horizon approaches the size $\mathcal{L}$. These interesting problems are currently poorly understood (see, for example, the paper [54], in which results different from those of [53] were obtained).

If the application of the AdS/CFT correspondence to black holes is justified and the corresponding hypotheses are valid, the existence of stellar-mass black holes itself imposes extremely strong constraints on the value of the fundamental AdS radius, namely, $\mathcal{L} < 10^{-3} - 10^{-4}$ mm (for the model of [53]), while the modern laboratory constraints are $\mathcal{L} \lesssim 0.1$ mm [55].

## 6. INITIAL BLACK-HOLE MASS FUNCTION: THE INVERSE PROBLEM

In spite of the hypothetical nature of the above concepts (starting from the adequacy of describing the Universe using models with macroscopic additional dimensions!), let us try to use them to explain the observed spectrum of the dynamically-measured masses of black holes in close binary systems. Namely, *let us suppose that the observed absence of black holes with masses below $4M_\odot$ is due to their rapid evaporation in the RS model.* Consequently, black holes with smaller masses cannot be observed, at least in old close binary systems. Of course, the collapse of a massive stellar core at the end of its evolution can give birth to a black hole with an even lower mass, but its lifetime will be short due to the enhanced evaporation. We emphasize again that the evaporation of black holes in this model occurs into unobservable CFT modes; i.e., from the viewpoint of a distant observer, the mass of the black hole decreases without any other detectable effects. The contribution of the possible evaporation of stellar-mass black holes to the total energy budget of the Galaxy is also negligible. Let us adopt the extreme assumption that all black holes formed via the evolution of ordinary stars over the Hubble time have evaporated. For our estimates, we take the average rate of star formation from baryons in the Galaxy to be $\sim 1 M_\odot/\mathrm{yr}$ and the lower limit for the initial stellar mass that can give birth to a black hole at the end of its evolution to be $30 M_\odot$. Then, for a Salpeter initial mass function, the mass of baryons transformed into black holes over the Hubble time should be about 1% of the total baryonic mass of the Galaxy.

Let us estimate the initial black-hole mass function $f_0(M)$ that is required to satisfy the observed black-hole mass distribution $f(M) = dN/dM \approx$ const for a given mass-variation law $dM/dt$. In the stationary case, the evolution of the one-dimensional distribution function is described by the kinetic equation

$$\frac{\partial}{\partial M}\left[f(M)\dot{M}\right] = f_0(M), \qquad (7)$$

which, in the case $\dot{M} < 0$ (evaporation), reduces to

$$f(M) = \frac{\int_M^{M_{\max}} f_0(M')dM'}{\dot{M}}, \qquad (8)$$
$$M > M_{\min}.$$

When $M \leq M_{\min}$, the form of the stationary distribution does not depend on the initial mass function, and is determined only by the black-hole mass-variation law:

$$f(M) = \frac{\int_{M_{\min}}^{M_{\max}} f_0(M')dM'}{\dot{M}} = \frac{\text{const}}{\dot{M}}, \qquad (9)$$
$$M \leq M_{\min}.$$

If the rate of evaporation is higher than the rate of increase in the mass (in close binary systems with an average accretion rate of $\sim 10^{-10} M_\odot$/yr, this condition is satisfied when $\mathcal{L} \gtrsim 10^{-2}$ mm for the model of [53]), then $dM/dt = \dot{M}^- \propto M^{-2}$. Assuming a power-law form for the initial black-hole mass function $f_0(M) \propto M^{-\alpha_i}$, we obtain $f(M) \sim M^{-\alpha_i+3}$ at $M > M_{\min}$ and $f(M) \sim M^2$ when $M \leq M_{\min}$, as is illustrated in Fig. 2 (we assumed $M \ll M_{\max}$ in the above estimates). It is interesting that a flat distribution is obtained when the coefficient of the slope of the initial black-hole mass function is $\alpha_i \sim 3$, close in absolute value to the slopes of the initial mass function of main-sequence stars ($\alpha_i \approx -2.5$) and the mass function of black holes in massive close binaries derived from observations of ultrabright X-ray sources in other galaxies ($\alpha_i \approx --2\ldots -2.2$).

The self-consistency of the hypothesis being considered can be tested as follows. The condition for the evaporation of a black hole with a mass below $M_0$ over the Hubble time in the model of [53] leads to a constraint on the AdS radius:

$$\left(\frac{\mathcal{L}}{1[\text{mm}]}\right)^2 \gtrsim 10^{-8}\left(\frac{M_0}{M_\odot}\right)^3. \qquad (10)$$

Consequently, cancelling out the factor $\mathcal{L}^2$ in the expression for the evaporation rate, we obtain

$$\left(\frac{dM}{dt}\right)^- \gtrsim 3 \times 10^{-11} M_\odot/\text{yr} \left(\frac{M_0}{M_\odot}\right). \qquad (11)$$

Therefore, the conditions for evaporation over the Hubble time and for an excess of the evaporation rate over the average accretion rate in close binary systems are satisfied simultaneously when $M_0 \gtrsim 10 M_\odot$. The value $M_0 \sim 4 M_\odot$ adopted in our analysis is lower than but fairly close to this limit (given the considerable model uncertainty in the numerical coefficients in the formulas describing evaporation). On the other hand, fixing the value $M_0 = 4M_\odot$ yields $\mathcal{L} \gtrsim 5 \times 10^{-4}$ mm and $\dot{M}^- \gtrsim 10^{-10} M_\odot$/yr, which are also consistent with the available constraints on $\mathcal{L}$ and the hypothesized decrease in the masses of black holes in close binaries due to their evaporation.

## 7. CONCLUSION

Analysis of the observed distribution of masses of relativistic objects (neutron stars and black holes) in close binary systems leads to the conclusion that the masses of neutron stars and black holes are distributed according to substantially different laws. The neutron-star masses are concentrated within the narrow range 1–2 $M_\odot$, while the black-hole masses are spread over the broad interval 4–15$M_\odot$, without a concentration near any specific mass. The uncertainties in the dynamical masses of black holes are due primarily to systematic errors introduced by the methods used to estimate the orbital inclinations and the component-mass ratios of the close binaries, derived from the light curves and spectra of the optical stars (related to the model dependence of the contributions of gaseous structures to the total optical luminosity of the systems). These uncertainties can be reduced by using refined models for the orbital variability of the absorption-line profiles in the spectra of the optical stars [12, 16, 17] and by using high-resolution spectra ($R = 50\,000-100\,000$) obtained with large modern telescopes when comparing the modeled and observed profiles. More accurate distances to X-ray binaries measured by next-generation space astrometric observatories (such as SIMA, GAIA, etc.) will also facilitate determinations of the orbital inclinations $i$ of the binary systems. It will also be useful to accumulate information about the masses of isolated black holes via observations of gravitational microlensing events.

The mass function of black holes in massive close binaries can also be derived from observations of ultrabright ($L_x > 2 \times 10^{39}$ erg/s) X-ray sources in other galaxies [37]. Assuming that these sources represent massive X-ray binaries radiating at the Eddington luminosity, the observed slope of the luminosity function at luminosities of $2 \times 10^{39}-2 \times 10^{40}$ erg/s leads to a power-law black-hole mass function $dN/dM \sim M^{-2.2}$, in contradiction with the lack of concentration of the black-hole masses near the lower limit $4M_\odot$, as follows from the derived dynamical masses of black holes in close binaries (most of which are low-mass systems).

The characteristic features of the black-hole mass distributions in these two cases can be reconciled under the hypothesis [52, 53] that the evaporation of black holes is enhanced on the RS2 brane due to the large number of (unobservable) CFT modes that appear in the extrapolation of AdS/CFT correspondence to black holes on the brane. This model can also explain the absence of observed black holes with masses $< 4M_\odot$ in low-mass close binaries with low average accretion rates. This hypothesis can be verified by searching for black holes with lower masses (both isolated and in binary systems).

Thus, the reliable determination of the mass function of compact relativistic objects in close binary systems is a very important observational problem of modern astrophysics. This function can be used both to test the general relativistic theory of the formation of neutron stars and black holes during the collapse of the cores of massive stars and to verify theories of gravity that are fundamental in a deeper sense.

## ACKNOWLEDGEMENTS

We are grateful to A.A. Starobinsky and D.V. Galtsov for fruitful discussions and comments. KAP was supported by the Russian Foundation for Basic Research (projects codes 02-02-16500, 03-02-16110, and 03-02-16068). AMC was supported by the Russian Foundation for Basic Research (project code 02-02-17524) and the Program of Support of Leading Scientific Schools of Russia (project 00-15-96-553).